Superlinear scaling in the urban system of England of Wales. A comparison with US cities

Erez Hatna

Center for Advanced Modelling in the Social, Behavioral and Health Sciences, Johns Hopkins University, USA

Abstract

According to the theory of urban scaling, urban indicators scale with city size in a predictable fashion. In particular, indicators of social and economic productivity are expected to have a superlinear relation. This behavior was verified for many urban systems, but recent findings suggest that this pattern may not be valid for England and Wales (E&W), where income has a linear relation with city size. This finding raises the question of whether the cities of E&W exhibit any superlinear relation with respect to quantities such as the level of education and occupational groups. In this paper, we evaluate the scaling of educational and occupational groups of E&W to see if we can detect superlinear relations in the number of educated and better-paid persons. As E&W may be unique in its linear scaling of income, we complement our analysis by comparing it to the urban system of the United States (US), a country for which superlinear scaling of income has already been demonstrated. To make the two urban systems comparable, we define the urban systems of both countries using the same method and test the sensitivity of our results to changes in the boundaries of cities. We find that cities of E&W exhibit patterns of superlinear scaling with respect to education and certain categories of better-paid occupations. However, the tendency of such groups to have superlinear scaling seems to be more consistent in the US. We show that while the educational and occupational distributions of US cities can partly explain the superlinear scaling of earnings, the distribution leads to a linear scaling of earnings in E&W.

## 1.0 Introduction

According to the theory of urban scaling (Bettencourt and West, 2010), urban indicators tend to scale with city size in a predictable manner. City size is usually represented as the number of residents, and the scaling relationship is formulated as a power law:

$$y = \alpha x^\beta \qquad (1)$$

where y represents an urban indicator and x the size of the settlement, α is a normalization coefficient, and β is the scaling exponent. Bettencourt et al. (2007) classify urban indicators into three categories:

(a) Indicators that scale proportionally with size (β ~ 1), which are related to basic human needs such as the consumption of electricity or the number of residential units.
(b) Indicators that scale sublinearly with size (β < 1), which exhibit decreasing returns to scale and include variables such as the total area of roads or the number of individuals working in the service sector.
(c) Indicators that exhibit superlinear scaling (β > 1), which indicate increasing returns to scale. These indicators are considered to be related to the productivity of cities and include financial services, patents, scientific products, gross domestic product, and income (Bettencourt et al., 2007).

Several mechanisms were proposed to explain the superlinear regularities, including more efficient sharing of resources in larger cities, better matching between different entities such as between employers and employees and between buyers and suppliers, better learning in larger markets, and stiffer competition that only allows the survival of the most productive firms ("firm selection") (Puga, 2010). One of the indicators for the productivity of a city is total income and wages. Income is considered as an indicator of productivity because it is believed to reflect the marginal product of labor (Glaeser, 2010). Accordingly, a superlinear scaling of income or wages may indicate that larger cities are more productive.

The tendency of the average pay to increase with city size is also related to the sorting of workers according to skill. Larger cities with advanced industries are likely to attract and maintain a greater share of the more educated, skilled, and better-paid workers compared to smaller and less successful cities (Eeckhout, 2004). Thus, the scaling relation between the number of skilled workers and city size should be superlinear, while a sublinear relation is expected for less skilled workers. This principle is demonstrated in a study of the level of innovation of United States (US) metropolitan areas. The superlinear relationship between the number of patents produced in US metropolitan areas and their population size was found to be due to the higher proportions of innovators and not due to a higher productivity of innovators in larger cities (Bettencourt et al., 2007b).

The evidence of increased productivity of larger cities implies that cities can mitigate the negative effects of size. For example, Angel and Blei (2015) have found that for US metropolitan areas, commuting time increases less than would be expected according to their size. They show that this propensity is related to three mechanisms that take place as cities grow: an increase in residential density, locational adjustments of residents and workplaces, and an increase in commuting speeds.

It is claimed that urban scaling laws should be valid for any urban system (Bettencourt, 2013). However, a recent study reveals that the urban system of England of Wales (E&W) may not fit the expected scaling portrait (Arcaute et al., 2015). The study suggests that residential income scales proportionally with city size, while it should exhibit a superlinear relation according to the scaling hypothesis (Bettencourt et al., 2007). This finding raises the question of whether the urban system of E&W exhibits any scaling patterns that are consistent with the theory and how that portrait differs from the scaling patterns of a country with a superlinear scaling of income.

In this paper, we evaluate the scaling of educational and occupational groups of E&W to see whether we can detect superlinear relationships. We are interested in seeing whether the number of skilled workers scales superlinearly even though the scaling of income is linear. As E&W may be unique in its linear scaling of income, we complement our analysis by comparing E&W to the urban system of the United States (US), a country for which superlinear scaling of income was already demonstrated (Bettencourt et al., 2007; Bettencourt et al., 2007b). To be more consistent in our analysis, we use the same method to define cities in the two countries. The method allows us to assess the sensitivity of the scaling exponents to variation in the boundaries of cities. To get a reliable estimate of the exponents, we use the method suggested by Leitao et al. (2016) alongside the Ordinary Least Squares (OLS) approach.

The paper has the following structure: Section 2 describes the data and a procedure for aggregating small statistical areas into cities. The results are presented in Section 3. We first describe the outcomes of the clustering procedure (3.1) and describe the scaling of income in the two countries (3.2). We then examine the scaling behavior according to education (3.3), industry of employment (3.4), and occupation group (3.5). In Section 3.6, we evaluate the sensitivity of the exponents to the way in which cities are defined and relate the scaling patterns of employment and occupation with the scaling of earnings (3.7). In Section

3.8, we briefly mention how the two regions countries differ with respect to the spatial autocorrelation of income, and we conclude the paper with a discussion (4.0).

## 2.0 Data and Methods

Data for E&W are taken from the United Kingdom (UK) census of 2001. We limit our study to E&W rather than the entire UK because the census of Scotland is incompatible with the census of E&W. Data for the US are taken from the US census of 2000. We limit our study to continental US by removing areas such as Alaska and Puerto Rico. We define cities based on small geographical units: census *wards* of E&W and census *tracts* of the US. We base our study on the previous censuses instead of the recent ones because the data availability of 2000 and 2001 censuses allows us to conduct the analysis for both countries in a similar fashion.

Census wards are small geographical units that are produced by the UK Office for National Statistics. Ward boundaries reflect the political geography of the UK at a fine resolution and, due to the need to maintain equality of representation in political elections, have similar populations. In this study, we use the wards defined in 2003. The region of E&W consists of 8,850 wards, the average number of residents within a ward is about 5,880, and the density is 20.2 persons per hectare.

US census tracts are small geographic units that are produced by the US Census Bureau. Similarly to wards, many variables of the census are given at this level. We use tracts that were defined for the 2000 census. Within our study area, the US is partitioned into 64,882 tracts. The average number of residents within a tract is about 4,308, and the average density is 20.3 persons per hectare.

Additional information on the data is available in the Appendix (Sections 6.3 and 6.4).

### 2.1. Aggregating small geographical units into settlements

We construct settlements and cities by aggregating wards (E&W) and tracts (US) into clusters. The clusters are formed based on the population density of each geographical unit. We first select all the units with population densities above a density threshold $d$ and then fill holes in the clusters. Finally, we assign each unit an identifier that indicates the cluster to which it belongs. Any pair of geographical units that are $m$ meters apart are assigned to the same cluster (see Appendix, Section 6.6.1).

Different values of the density threshold $d$ produce different urban systems. High-density thresholds produce small urban cores while low thresholds lead to large clusters. We use a simple quantitative criterion for setting $d$ whereas we set $m$ manually.

To select a density threshold that produces a realistic system of cities, we compare each realization of clusters with a land-cover map. We use the Corine 2000 raster (EEA, 2002) for E&W and the National Land Cover of 2001 (Homer et al., 2001) for the US. We do so by calculating the Person Product Moment Coefficient between the set of geographical units composing the clusters and the land-use categories (see Section 6.6.4 in the Appendix).

To investigate the sensitivity of our results to the way in which the urban system is defined, we increase the number of systems defined by taking into account the flow of commuters into each cluster. This step can produce clusters that better represent functional areas such as metropolitan areas. In our method, a ward or a tract is assigned to a cluster if at least a percentage $F$ of its working residents are commuting to that cluster (see Section 6.6.2 of the Appendix). Commuting data for the US were taken from the "Census

2000 Special Tabulation: Census Tract of Work by Census Tract of Residence (STP 64)" table, while for E&W, we used the "2001 Census: Special Workplace Statistics" table.

## 2.2. Estimating the scaling exponents

According to (1), we consider the expectation of an urban indicator to be conditioned on city size as follows: $E(y|x) = \alpha x^\beta$. The scaling exponent β is estimated using two approaches. The first approach is the classic OLS method using the logarithms of x and y. We evaluate the null hypothesis that β = 1 by constructing a 95% confidence interval around β. The calculations were performed using the *lm* function of the R software (R Core Team, 2013).

The second approach was formulated by Leitao et al. (2016). The method is a probabilistic formulation of the scaling law that makes it possible to test the hypothesis that β = 1 by avoiding many of the limitations of the least squares approach. Leitao et al. (2016) estimate the expectation of an urban indicator conditioned on city size as a power law and specify a conditional variance using Taylor law: $Var(y|x) = \gamma E(y|x)^\delta$. The distribution y|x is defined as either Gaussian or log-normal. We compare the two models using the Bayesian Information Criterion (BIC) and present the better one. We mark an exponent estimated by the Gaussian model with "G" and an exponent estimated by the log-normal model with "L". In all our calculations, we use the models that includes δ as a free parameter.

To infer whether a scaling exponent is linear or nonlinear, we calculate a corresponding model where β = 1 and take the difference in the models' BICs ($\Delta BIC = BIC_{\beta=1} - BIC_\beta$). If $\Delta BIC$ < 0, we conclude that the relation is linear; if $\Delta BIC \geq 6$, we conclude that the relation is nonlinear. We categorize the rest of the cases as inconclusive. We mark linear exponents with "~" and non-linear ones with "*". The method allows us to reject the model; we mark exponents that are not rejected with "^". According to Leitao et al. (2016), most relations tend to be rejected because they do not strictly follow the model assumptions. For example, a non-linear exponent estimated using the Gaussian model would be marked as "G*" if the model is rejected or as "G*^" if the model is not rejected.

We conduct the analysis using the Python code that was kindly provided by Leitao et al. (2016). We refer to exponents derived using this method as maximum likelihood estimations.

# 3.0 Results

## 3.1. Defining the urban systems

The best density for E&W is 17 persons per hectare (Figure 1). We use a distance threshold of 1 kilometer, which produces 453 clusters ranging in population size from 1,316 to 7.4 million inhabitants. Under these settings, cities are not broken into smaller pieces, while nearby cities, like Birmingham and Liverpool, are represented by separate clusters.

For the US, we observe the highest correlation at d = 6.5 persons per hectare. We use a larger merging distance of 2 kilometers because we would like to keep multicentric cities as single entities. Baltimore and Washington tend to merge at *d* = 6.5. To keep these two cities separated, we set the density at *d* = 8 persons per hectare. This value is sufficiently close to the maximum while preserving the two cities. Under these settings, we get 1,500 clusters with populations ranging from 555 to 15.6 million inhabitants.

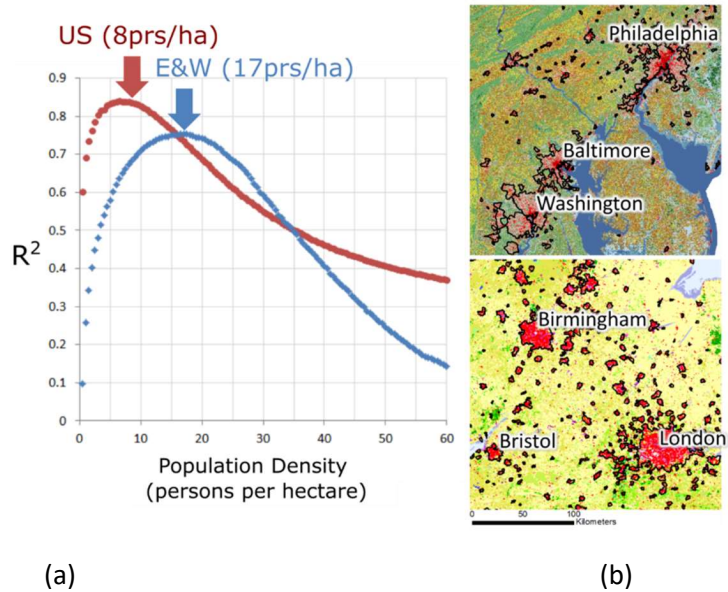

(a)                          (b)

Figure 1: The correspondence between clusters and land-use maps for E&W and the US: (a) the correlation for various density thresholds (measured in persons per hectare); (b) maps depicting the best fit for cities near London and Washington.

## 3.2 The scaling of income

The scatterplots of total income versus the number of residents based on the chosen densities are shown in Figure 2. For the US, the relationship is superlinear ($\beta = 1.04$), while for E&W, the relation is linear. Thus, total income seems to be proportional to size in E&W. The model underestimates London, the largest city by far. The linear relationship between total income and city size demands further validation since this outcome does not follow the expected superlinear behavior and because the income data are model based. Additional evidence based on the 2012 census and NUTS3 regions supports the linear relationship as well (see Sections 6.1 and 6.2 of the Appendix).

For the US, the census provides data on the number of inhabitants within different ranges of income (Figure 2, bottom). As expected, we find the exponents of lower incomes (below $35,000 a year) to be sublinear while higher incomes (above $50,000 a year) are superlinear. A similar pattern was found in Australia by Sarkar et al. (2016). The OLS estimations for lower income ranges differ from the maximum likelihood estimations, which we consider to be more reliable.

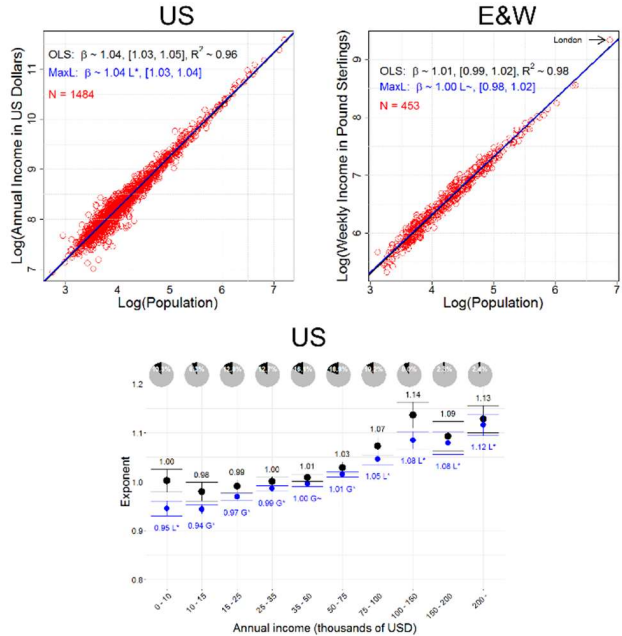

Figure 2: The scaling of total income in relation to total population for the US and E&W based on *d* = 8 and 18 persons per hectare respectively (top), and the scaling exponents for the number of people with various income ranges for the US (bottom).

### 3.3. The scaling of level of education

An urban system characterized by superlinear scaling of income is likely to have a superlinear scaling of the number of better-educated persons and a sublinear scaling for the number of less-educated ones. The scaling exponents for the number of adults with a given level of education in the US are shown in Figure 3. City size is defined as the total number of adults (instead of total population) to reduce the potential effect of age differences in cities (see Section 6.5.1 for a description of the education variables). The scaling of the number of persons with high-school education is sublinear ($\beta$ = 0.95), while the scaling is superlinear for higher education levels such the number of individuals with a master's degree ($\beta$ = 1.05). Exceptions to the expected pattern of scaling include the No Education category, which scales linearly, and the linear scaling of the number of people with a doctorate. However, these are small categories, each representing less than 2% of the population in our system of cities, and they do not affect the general upward trend of the exponents (Figure 3b). Together, the superlinear categories amount to about 47% of the adult population is our system of cities. Estimated exponents of the OLS method are higher than the ones of the maximum likelihood method, which in some cases affects the classification of the exponents.

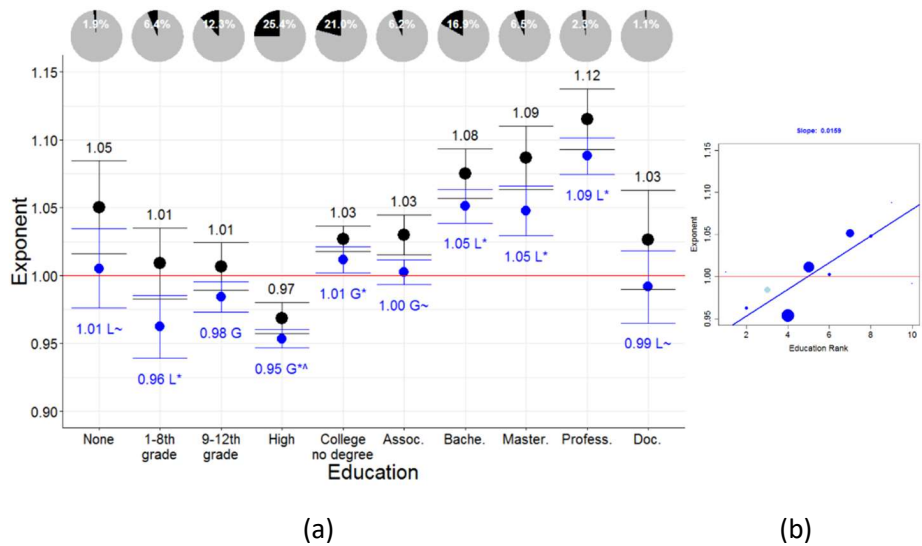

(a) (b)

Figure 3: (a) Scaling exponents for the highest level of education completed by an individual in the US in 2000. City size is defined as the number of adults (25 years and over). The categories are ordered from lower (left) to higher education levels (right). Pie charts depict the relative size of the educational cohort out of the total adult population. (b) The exponents versus the education rank. The size of the points is proportional to the size of the education cohort. The blue trend line is estimated using the weighted least squares method (the weight is the size of the cohort).

Figure 4 reports the scaling exponents of education for E&W. The UK census aggregates education into five qualification levels (Section 6.4.2 in the Appendix). The scaling exponents are sublinear (β = 0.98) for lower secondary school qualifications (levels 1 and 2) while the exponent of the higher secondary school qualification category (level 3) is superlinear (β = 1.06). However, a linear relation exists for the no-qualification category and also for the highest education level (levels 4/5), which represent academic degrees. Contrary to the maximum likelihood method, the OLS estimation indicates a superlinear relation for the highest education level.

Thus, we find that the US has a pattern of scaling that indicates a larger fraction of educated adults in larger cities: lower education levels are sublinear while higher levels are superlinear. For E&W, we find a superlinear relation for the highest secondary school qualification, but the relation with respect to academic degrees is linear.

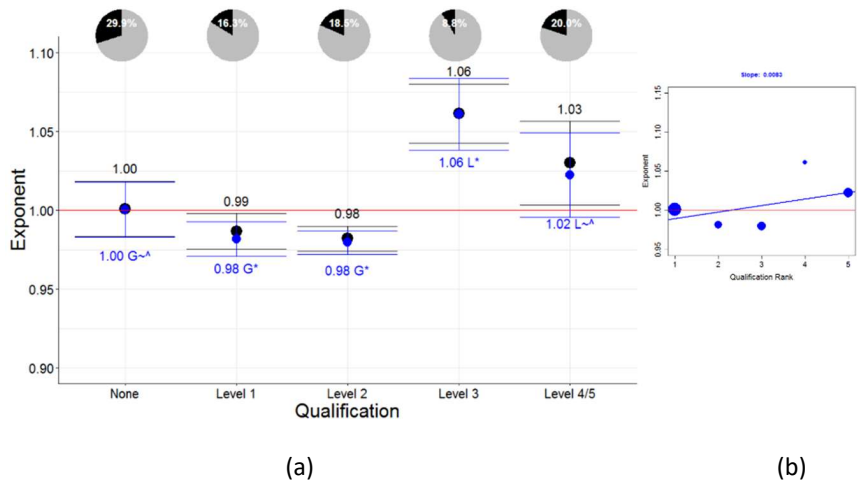

(a)                          (b)

Figure 4: (a) Scaling exponents for qualification levels for E&W in 2001. City size is defined as the number of adults (16 to 74 years). The categories are ordered from lower (left) to higher qualification (right). The pie charts depict the relative sizes of the qualification cohorts out of the total adult population. (b) The exponents versus the qualification rank. The size of a point is proportional to the size of the qualification cohort. The blue trend line is estimated using the weighted least squares method (the weight is the size of the cohort).

### 3.4. The Scaling of Industry of Occupation

To gain further insight, we examine how the number of workers employed in different industries scales with city size (for information on these variables, see Sections 6.4.3 and 6.2.2). Figure 5a presents the scaling exponents of US industry of occupation categories. The industries are shown in ascending order from left to right according to the national average pay of each category. The exponents are plotted against the average pay of the category in Figure 5b. We represent city size as the total number of workers to reduce the potential effect of different age structures and unemployment rates. We expect to find a pattern where the number of workers in lower paid industries tends to scale sublinearly, while a superlinear relation would be likely for better-paid industries.

We find that the exponents for the number of workers in highly paid industries are indeed superlinear, while the lower-paid manufacturing cohort scales sublinearly. However, the exponents are mostly linear for industries with an average pay below that of manufacturing ($44,000 a year). We find a clear pattern of superlinear exponents of the four highest paid categories, which is probably related to the much higher average pay of these categories (Figure 5b). Together, the superlinear categories account for about 26% of the working population of the system of cities we use.

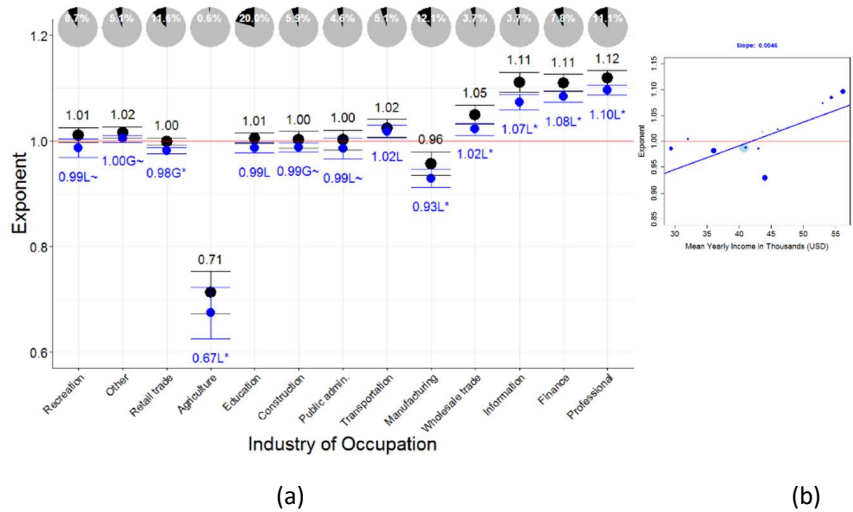

Figure 5: (a) Scaling exponent for the number of adults (16 years and over) working in different industries in the US. City size is defined as the total number of employed adults. The industries are ordered according to national average pay from low (left) to high (right). The pie charts depict the fraction of individuals employed in each industry out of the total employed adult population. (b) The exponents versus the average pay.

The scaling exponents for the industry of occupation in E&W are shown in Figure 6a. The category of financial intermediation, which has the largest average pay by far, has a superlinear exponent of 1.09. This category contains about 5% of the population of our system. The two other superlinear industries are real-estate and transport (β = 1.06 for both). Together, the superlinear categories account for about 26% of the working population in our system of cities. All the categories with average pay below that of transport are linear, except for agriculture. Thus, we find the that the better-paid categories tend to have superlinear exponents in E&W (Figure 6b), but this pattern is less clear than the one of the US in terms of the number of superlinear categories and the smaller magnitude of the superlinear exponents. In both countries, many of the lower-paid categories tend to be linear. This may be the case because the industry of occupation categories are crude categories with many sub-industries, each with a different scaling exponent and average pay. In both countries, a highly sublinear scaling exists for the number of individuals working in agriculture, but the relative number of such workers is small.

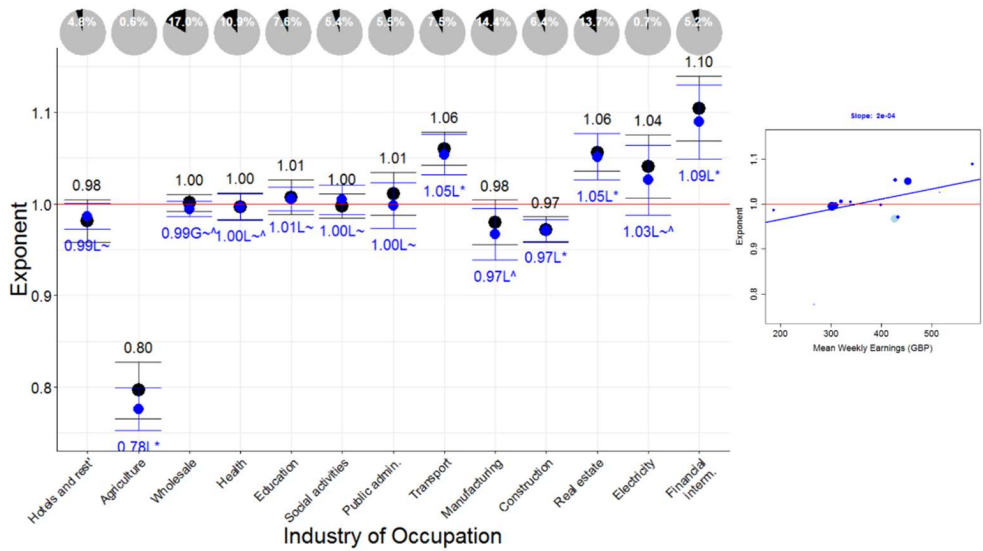

(a)                                                 (b)

Figure 6: (a) Scaling exponents for the number of adults (16 to 74 years) working in different industries in E&W. City size is defined as the total number of employed adults. The industries are ordered according to national average pay from low (left) to high (right). (b) The exponents versus the average pay.

### 3.5. The scaling of occupational groups

The censuses of both countries provide data on the occupation of workers. Occupation describes the kind of work a person does as a job (for more information, see Sections 6.4.4. and 6.5.3). For the US (Figure 7), we find that the management & professional category, which has a considerably higher pay compared to the other occupations (Figure 7b), also has the highest superlinear exponent ($\beta$ = 1.05). Two categories, construction & maintenance, and sales & office occupations, have similar average pay, but their scaling exponents are different: The first scales sublinearly ($\beta$ = 0.97) while the second scales superlinearly ($\beta$ = 1.02). This shows the limitation of using average pay to order categories. The rest of the occupations are sublinear as expected. Even though the pattern of scaling exponents is complicated, we do see a tendency for the better-paid occupations to have higher exponents (Figure 7b). The superlinear categories account for 53% of the working population in this system of cities.

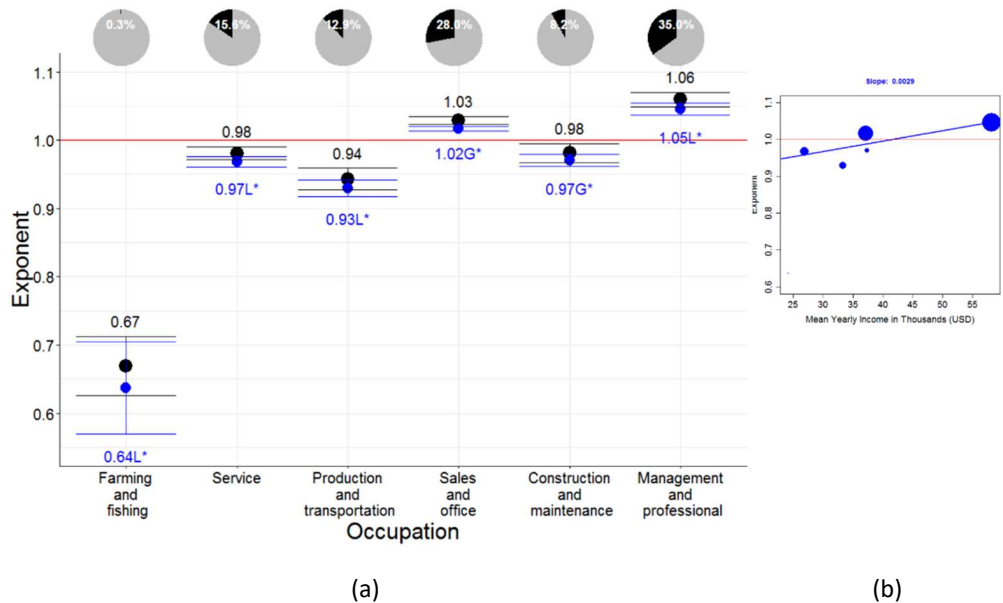

(a)                                        (b)

Figure 7: (a) Scaling exponent for the number of adults (16 years and over) working in different occupations in the US. City size is defined as the total number of employed adults. The employment categories are ordered by the national average pay from low (left) to high (right). (b) The exponents versus the average pay.

The scaling pattern of occupational groups of E&W is complicated as well (Figure 8). The exponents seem to be weakly dependent on average pay. The number of people employed in technical & associate professional occupations (β = 1.02), and professional occupations (β = 1.06) scale superlinearly as expected, but the highest-paid category of managers & senior officials does not.

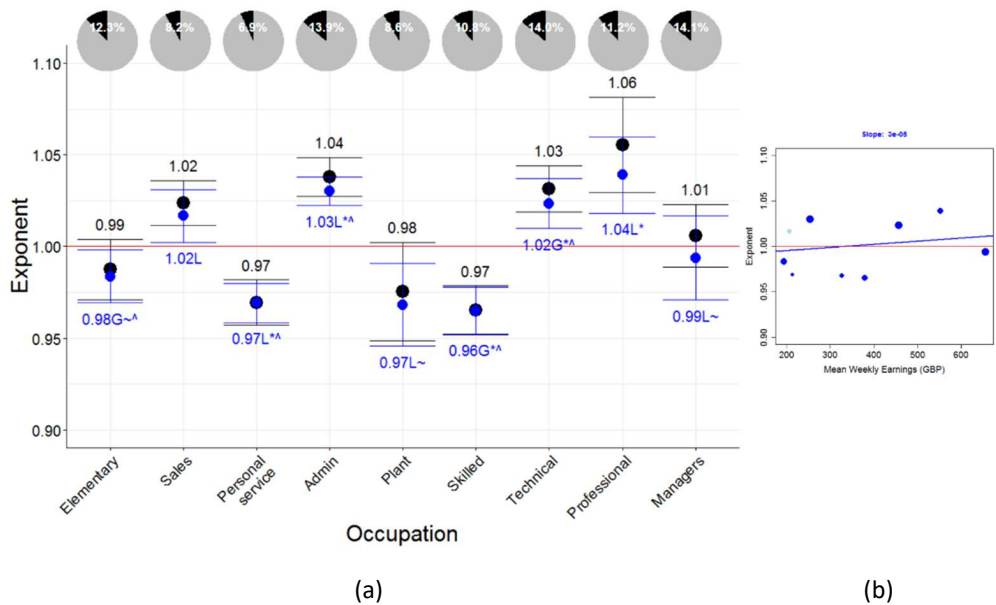

(a)                                        (b)

Figure 8: (a) Scaling exponents for the number of adults (16 to 74 years) working in different occupation categories in E&W. City size is defined as the total number of employed adults. The occupations are ordered by national average pay from low (left) to high (right). (b) The exponents versus the average pay.

Thus, we find superlinear scaling for several occupational groups of E&W, but many of them do not have a high average pay. This shows the limitation of using average pay for ordering the categories. For example, the superlinear scaling of the number of persons employed in Administrative roles is consistent with the notion of greater economic activity of larger cities, while the average pay in this group is relatively low. We found similar behavior for the US as well. In total, the superlinear categories account for about 24% of the cities' working population.

### 3.6 The sensitivity of the exponents to the definition of the urban system

To further evaluate our results, we test the sensitivity of the exponents to the way in which cities are defined. We calculate exponents for different population cutoffs that omit cities below a given size and for different combination of the population and commuting thresholds used in the clustering method (illustrated in Figure 9).

Figure 10a shows how the density and commuting threshold affects the scaling of income for the chosen densities. For the US, the superlinear relationship is preserved for different cutoffs, but the exponents increase from 1.04 to 1.06 as the cutoff increases from 0 to 30,000 inhabitants. For E&W, the linear relation is maintained regardless of the cutoff, and there is even a tendency for the exponent to decrease until a cutoff of 20,000 inhabitants is reached. The statistical model is not rejected for all investigated cutoffs above 0.

By estimating the exponents for different definitions of cities (Figure 10b), we find that for the US, the superlinear relation of income is preserved for investigated densities below 15 persons per hectare or for cases with low commuting thresholds (i.e. cases where many commuters are added to cities). It occurs because US cities tend to break into small cores at high densities. For E&W, the linear relation is insensitive to the two parameters.

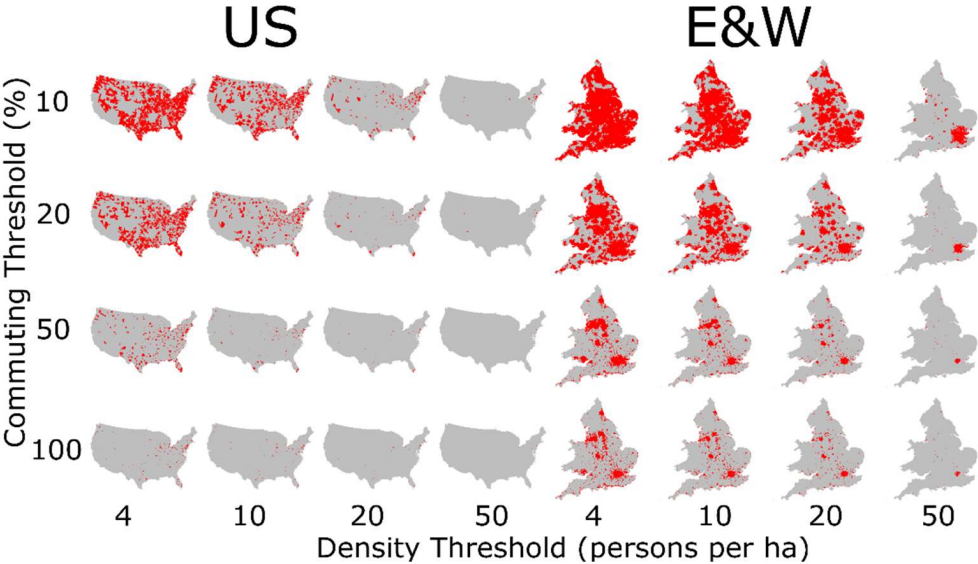

Figure 9: An illustration of 16 definitions of urban systems of the US and E&W based on the density and commuting thresholds.

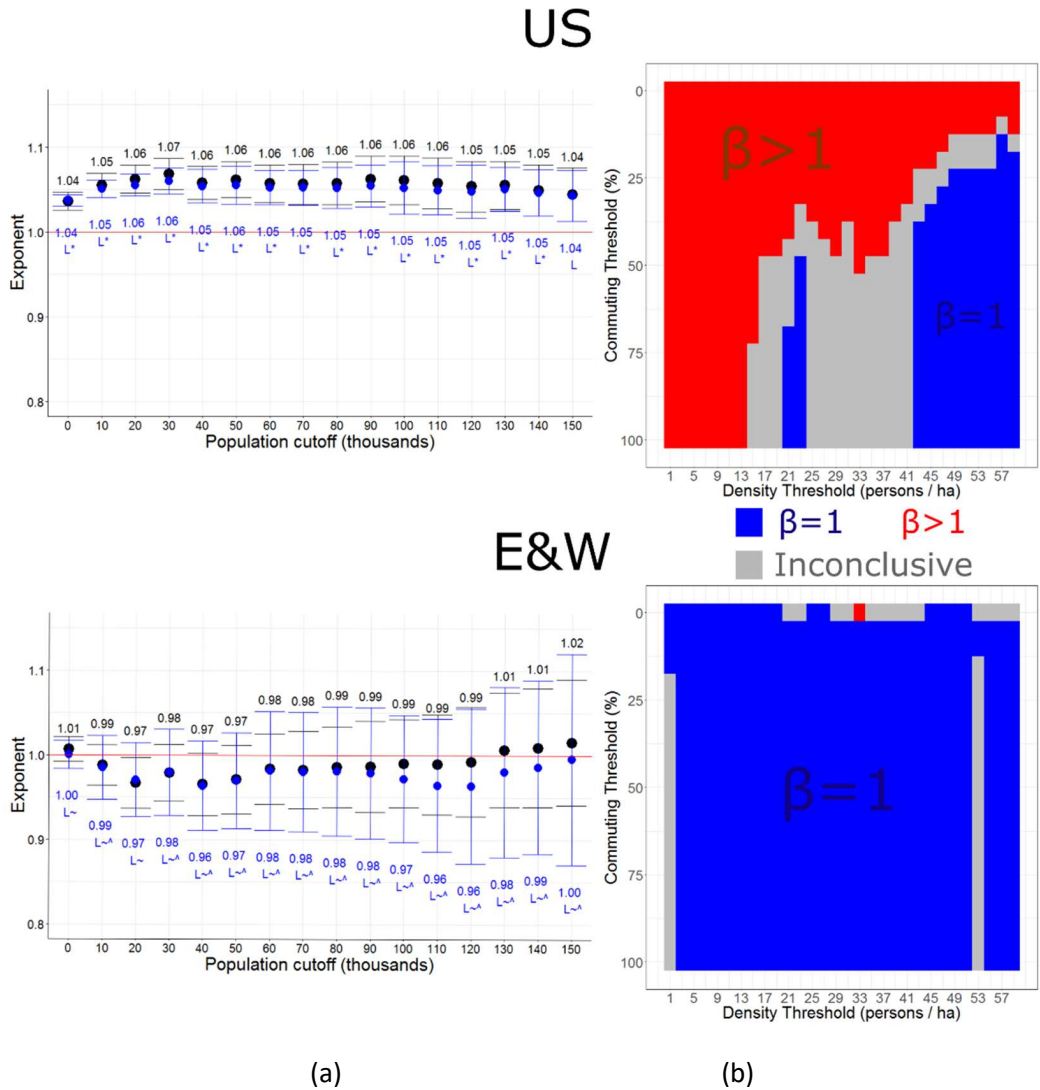

(a)                  (b)

Figure 10: The scaling exponents of income for the US and E&W as dependent on (a) population cutoffs and (b) the density and commuting thresholds. The regimes in the heatmaps are based on the log-normal model

We already found that the number of people employed in finance is superlinear in both countries. We now test the sensitivity of this indicator. (Figure 11). For the US, the superlinear relation is preserved for all investigated cutoffs, and we again find a moderate tendency of the exponents to increase from 1.08 to 1.11 for a cutoff between 0 and 30,00 inhabitants. For E&W, the superlinear relation is preserved for the two lower cutoffs (0 and 10,000 persons). The rest of the cutoffs (excluding 30,000 inhabitants) lead to either a linear or inconclusive result. This sensitivity to the cutoff is probably due to the smaller set of cities that remains at higher cutoffs. We find that the model is not rejected when cutoffs of 20,000 inhabitants and above are applied.

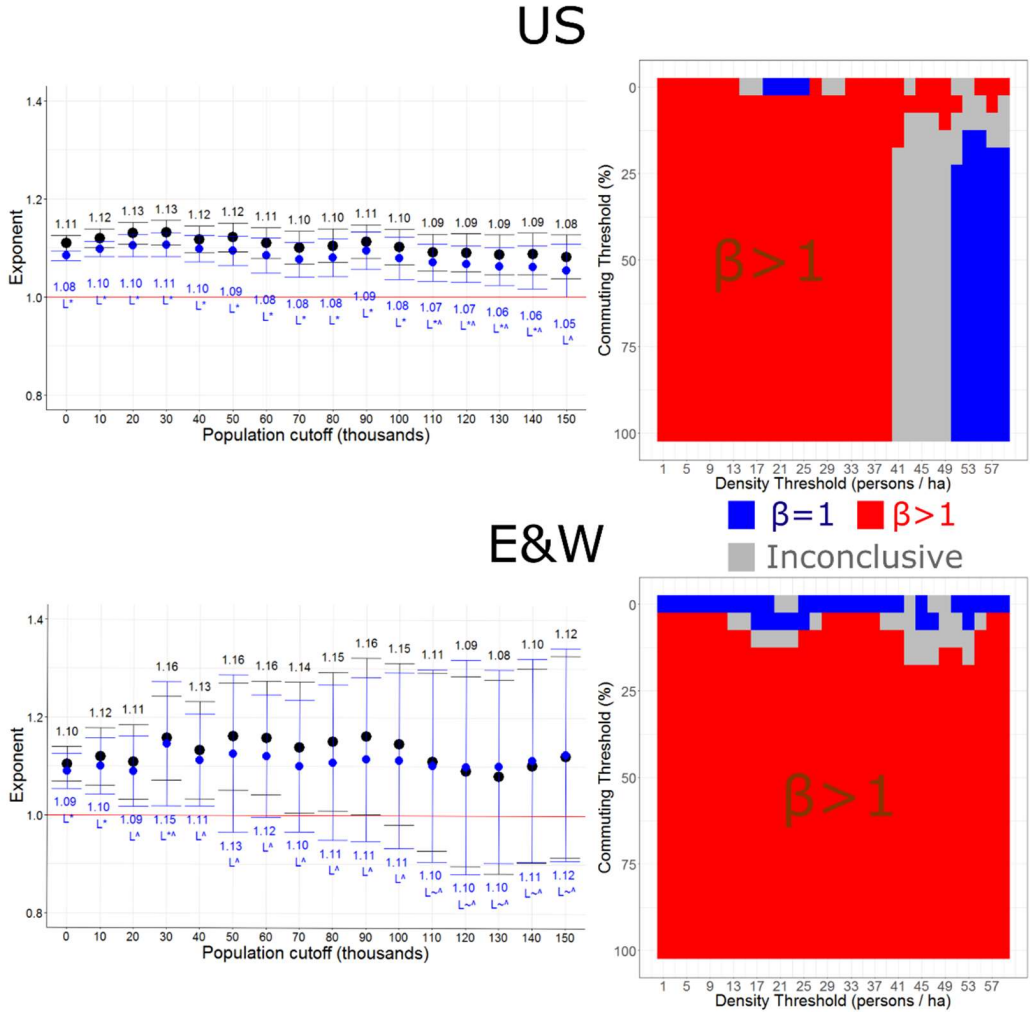

Figure 11: The scaling exponents of income for the US and E&W as dependent on (a) population cutoffs and (b) and the density and commuting thresholds. The regimes of the heatmaps are based on the log-normal model

Most of the studies of the US urban system make use of Core-Based Statistical Areas (CBSAs). To make sure our clustering method produces results that are compatible with these geographical units, we reproduced all the results based on CBSAs (Section 6.3 of the Appendix). The scaling exponent of income based on CBSA is 1.08 (Figure S3), whereas it is 1.04 in our clustering results. It seems that the population cutoff partly contributes to this because the CBSA system of cities is based on a cutoff of 10,000 inhabitants. For the same cutoff, the scaling based on our clusters is 1.05 (Figure 5), and it reaches higher values for higher cutoffs.

The pattern of scaling of education based on CBSA units (Figure S4) is similar to the one we found based on our clustering method. However, the scaling pattern is closer to our expectations: all education categories above the college levels are superlinear while the ones below (excluding No Education) are sublinear. The scaling exponents of industry of employment (Figure S5) and occupation categories (Figure S6) are also similar to the ones we already observed, but the superlinear exponents are larger.

We learned from this sensitivity analysis that for the US, our results are valid as long as the density cutoff used in the clustering method is not high, while for E&W, the results are valid as long as the population cutoff is low. We also found that the US scaling patterns are compatible with the ones based on CBSA.

## 3.7 Estimating the scaling of earnings

E&W has a linear scaling of income and a superlinear scaling of the number of better-educated persons and individuals working in better-paid industries. Are these two findings inconsistent?

To investigate how the number of individuals working in different industries and occupations affects the scaling exponents of income, we estimate the total earnings of each cluster according to the distribution of workers. We use earnings instead of income because data on earnings are available for occupational groups. We would like to see whether the estimated earnings scale linearly for E&W even though the number of better-paid individuals scales superlinearly. We also test whether the superlinear scaling of better-paid American workers can produce superlinear scaling of earnings.

We estimate earnings using two approaches: (1) by assuming that all workers earn the national average pay and (2) by assuming that all workers earn the national average pay of their occupational or industrial category. The scaling exponents of the first approach are equivalent to the scaling of the number of workers in relation to population. For the second model, we first need to estimate the total earnings in each cluster based on the number of workers and the national average pay of each category. By assuming that each worker is earning the national average salary of his or her category, we are likely to underestimate the earnings in large cities where workers' salaries are higher. For this reason, we expect to obtain scaling exponents that are closer to linear compared to the empirical ones.

The scaling exponents for the two models are shown in Figure 10. We use several population cutoffs to test the robustness of the estimations. For the US, using the first approach (which uses the same national average pay for all workers) we attain a superlinear exponent ($\beta \sim 1.01$) when no cutoff is applied and linear or inconclusive outcomes when cutoffs are used. Using the second approach, which uses the average pay of each group, we get a mild superlinear scaling (between 1.01 and 1.02) for all cutoffs up to 20,000 inhabitants for industry of occupation and 30,000 inhabitants for occupation. As expected, we obtained scaling exponents that are smaller than the one for income ($\beta \sim 1.04$), but we see that both the number of workers and the distribution of workers by category contribute to the superlinear relation. The exponents of the estimated earnings do not grow as the population cutoff increases like they do for income (Figure 5, left).

For E&W, we were not able to produce a superlinear scaling using this method. We find that the exponents of the two approaches are very similar, and both imply a linear relation. Thus, the higher concentration of better-paid categories of workers in larger cities does not produce superlinear scaling. We can see that the exponents of income tend to get smaller as the cutoff increases from 0 to 30,000. This is probably due to variation in the number of workers. We conclude that these results are consistent with the linear scaling of income.

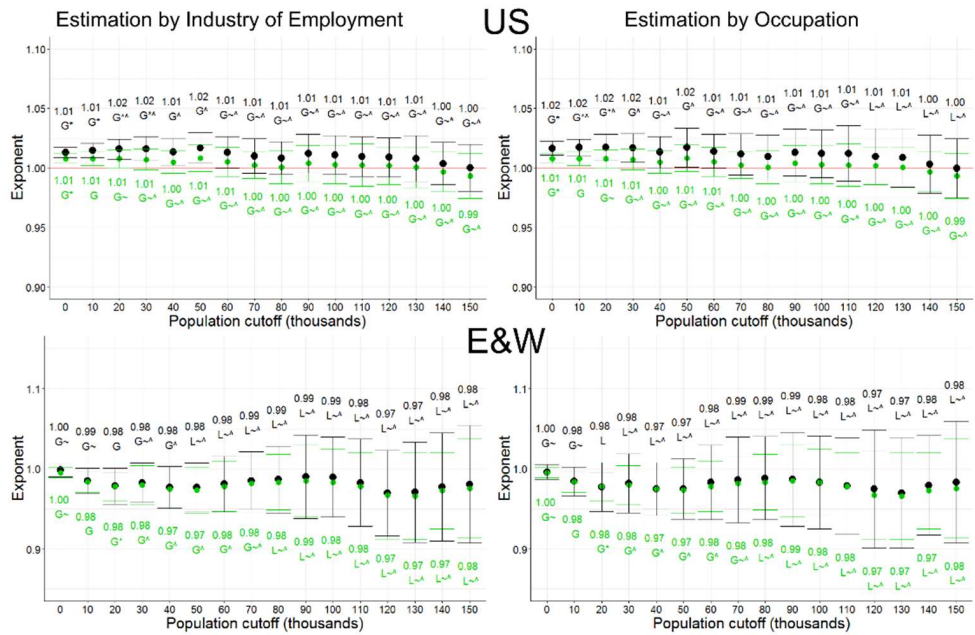

Figure 10: The scaling of estimated earnings in settlements in the US and E&W for different population cutoffs. Green points and intervals represent estimation based on the number of workers while black represents estimation based on the distribution of industry of employment (left) or the distribution of occupations (right)

### 3.8 Regional effects

Our analysis shows that the size of a city affects its total income in the US while this seems not be the case in E&W. One aspect of the difference between the two countries is apparent by simply looking at the geographical distribution of cities' average income (Figure 11). In both countries, the average income of cities is spatially correlated, but this tendency seems to be more striking in E&W, where the region around London has a higher average income compared to the rest of the country. This pattern is referred to as the North–South divide.

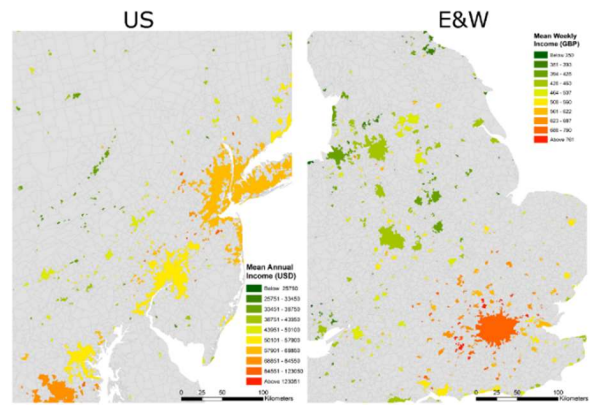

Figure 11: Average income of cities in E&W and the US.

To quantify this impression, we estimate the Moran's I index of spatial autocorrelation (Anselin, 1995) based on the mean income of cities for three distance thresholds (see Section 6.7 of the Appendix and Table 1). We find a significant autocorrelation in both countries, indicating that neighboring cities tend to have similar incomes. However, we also find that the index for E&W is indeed larger compared to the one for the US. This finding indicates that regional influences are more dominant in E&W, whereas in the US, both size and location contribute to cities' incomes.

| Distance Threshold | E&W | US |
| --- | --- | --- |
| 100 km | 0.79 | 0.61 |
| 200 km | 0.72 | 0.55 |
| 300 km | 0.69 | 0.52 |

Table 1: Moran's I of the average income of cities' residents for E&W and the US. All values are statistically different from 0 at a confidence level of 99%.

## 4.0 Discussion

The motivation for this research originates from the unusual scaling relation of income in the urban system of E&W, where the total residential income scales linearly with the number of residents. Other countries, such as the US, exhibit superlinear relationships. The linear relation of income and size raises the question of whether other quantities of E&W cities exhibit sublinear or superlinear scaling in a manner that is consistent with the theory of urban scaling. For this purpose, we investigated how the number of adults with varying levels of education and the number of workers employed in different industries and occupations scale with city size. In order to compare E&W with an urban system that exhibits a superlinear scaling of income, we performed the analysis for cities in the US as well. Instead of using predefined boundaries of cities, we defined cities in a consistent manner using the same clustering procedure in both countries. This procedure allowed us to test the sensitivity of the exponents to variation in the definition of cities. To better estimate the scaling exponents, we did not rely solely on the OLS method. Instead, we used the statistical model developed by Leitao et al. (2016).

We tested how the number of persons with different education levels scales with size. In the US, we found a clear pattern where education categories above high-school level tend to have superlinear exponents. For E&W, the number of persons with academic qualifications was found to scale linearly with size. However, we did observe a sublinear scaling for lower secondary school qualifications and superlinear scaling for higher-level ones. Thus, we observe a pattern of superlinear scaling of education in E&W but it is not as clear as the one in the US. This difference may be related to the different education categories that are provided by the censuses of the two countries.

In both countries, the number of workers employed in higher-paid industries scales superlinearly, but this pattern is clearer in the US. For the number of individuals working in different occupations, we found nonlinear exponents in both countries, but they seem to be weakly dependent on the average pay.

An evaluation of the sensitivity of the income exponent to variation in the definition of cities shows that for the US, the superlinear relation is valid if the density threshold is not high. We also found a tendency of the income exponent to grow when a population cutoff is applied. For E&W, we found that the linear relation between total income and number of residents is valid for a wide range of definitions. The exponents are also insensitive to population cutoffs. We repeated the sensitivity analysis for the category of financial intermediation, which showed a superlinear relation in E&W. We found that like total income, the scaling relation of the number of people working in financial intermediation is insensitive to the way in which cities are defined. Thus, it seems that the superlinear exponents of E&W are not an artifact of the way in which cities are defined. For the US, the finance category exhibits a similar dependency on density to total income.

Our results show that E&W exhibits patterns of superlinear scaling that are consistent with the notion that larger cities are more productive. This may seem inconsistent with the linear scaling of income. To check the consistency of our results, we estimated the sum of wages of all workers in each city by assuming that each worker earns the average wage of his or her group. We found that for the US, both the total number of workers and the relative fraction of more qualified workers in larger cities contribute to a superlinear scaling of earnings. In contrast, we found that the total number of workers scales linearly with city size in E&W and the higher proportions of skilled workers in larger cities do not produce a superlinear scaling of earnings. Thus, we found no inconsistencies in our results.

In both countries, the average residential income in cities is spatially autocorrelated, but this tendency is stronger in E&W. Cities near London tend to have a higher income than, for example, cities near Liverpool. Small and large cities that are close together tend to have similar average incomes, while in the US we do see differentiation by size. This could suggest that cities in E&W are more integrated with one another in a way that makes their physical boundaries less relevant for analysis. While the US contains several large cities, the UK, which is smaller, seems to be dominated by London. The existence of a primate city may not necessarily lead to a pattern of linear scaling. Evidence for superlinear scaling of wages and income was found for France, which is another country with a primate city (Cottineau et al., 2016). However, this study also showed that this superlinear relation might be sensitive to the way in which cities are defined.

For most of the variables, the maximum likelihood method (Leitao et al. 2016) produced exponents and confidence intervals that are similar to the corresponding OLS estimations. In general, conducting this research based on OLS estimation alone would not have altered our conclusions. In some cases, such as the scaling of categories of workers with low income in the US, the maximum likelihood method produces results that seem more consistent with the expected behavior. According to Leitao et al. (2016), the power law relation between city size and most urban indicators can be rejected using the models they tested. The outcomes of our study are consistent with this result, but we found that most of the cases where the model is not rejected occur for E&W, especially when a population cutoff is applied. The failure of the statistical model raises the need for new models that could better represent the relation between size and urban indicators.

Because the linear relation of income and size does not follow the expected pattern of scaling, we estimated the scaling exponents of income of E&W using the data of the censuses of 2000 and 2011 and based on NUTS 3 units and we checked the sensitivity of the exponents to variation in the boundaries of cities. We could not reject the hypothesis that the exponents are linear. However, further analysis of this relation may still be needed to verify these results. There is also a need to study the patterns of scaling of additional countries to determine whether E&W is unique in its scaling behavior and to characterize the properties that set E&W apart from other countries.

# 6.0 Appendix
## 6.1 Income in E&W by NUTS3 units

The linear relationship between total income and city size in E&W demands further validation because the results do not follow the expected superlinear relationship and because the ward level data are model-based estimations. For these reasons, we make use of the Regional Gross Disposable Household Income (GDHI) data that are published by the UK Office for National Statistics.

Household disposable income represents the amount of money that all of the individuals in the household sector have available for spending or saving after income distribution measures (for example, taxes, social contributions, and benefits) have taken effect. The GDHI estimates cover the UK as a whole and are broken down into 139 NUTS 3 (Nomenclature of Units for Territorial Statistics) that represent individual counties and unitary authorities. These estimates are based on place of residence; that is, the income of individuals is allocated to the region in which they live. The data were taken from Tables 3.1 and 3.2 of the NUTS 3 Regional GDHI, 1997–2011 dataset.

To construct the urban system of E&W based on NUTS 3 units, we create the best NUTS3 representation of density-based clusters (17 persons per hectare) that is used in the paper. This was achieved by overlaying the NUTS 3 units with the clusters. For each unit, we perform the following steps:

1. We detect the clusters with which the NUTS 3 unit intersects and calculate the area of each intersection.
2. We then detect the cluster with the largest intersection area.
3. If the fraction of the intersection out of the total area of the NUTS 3 unit is above 5%, the NUTS 3 is considered as urban, and the ID of the cluster is assigned to it. Otherwise, the unit is removed.

The method results in 40 cities, which are presented in Figure S1.

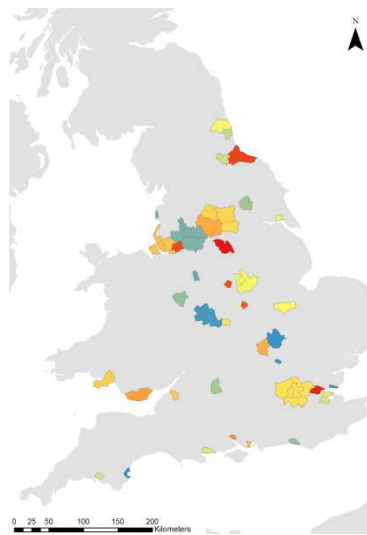

Figure S1: NUTS 3 representation of cities in E&W

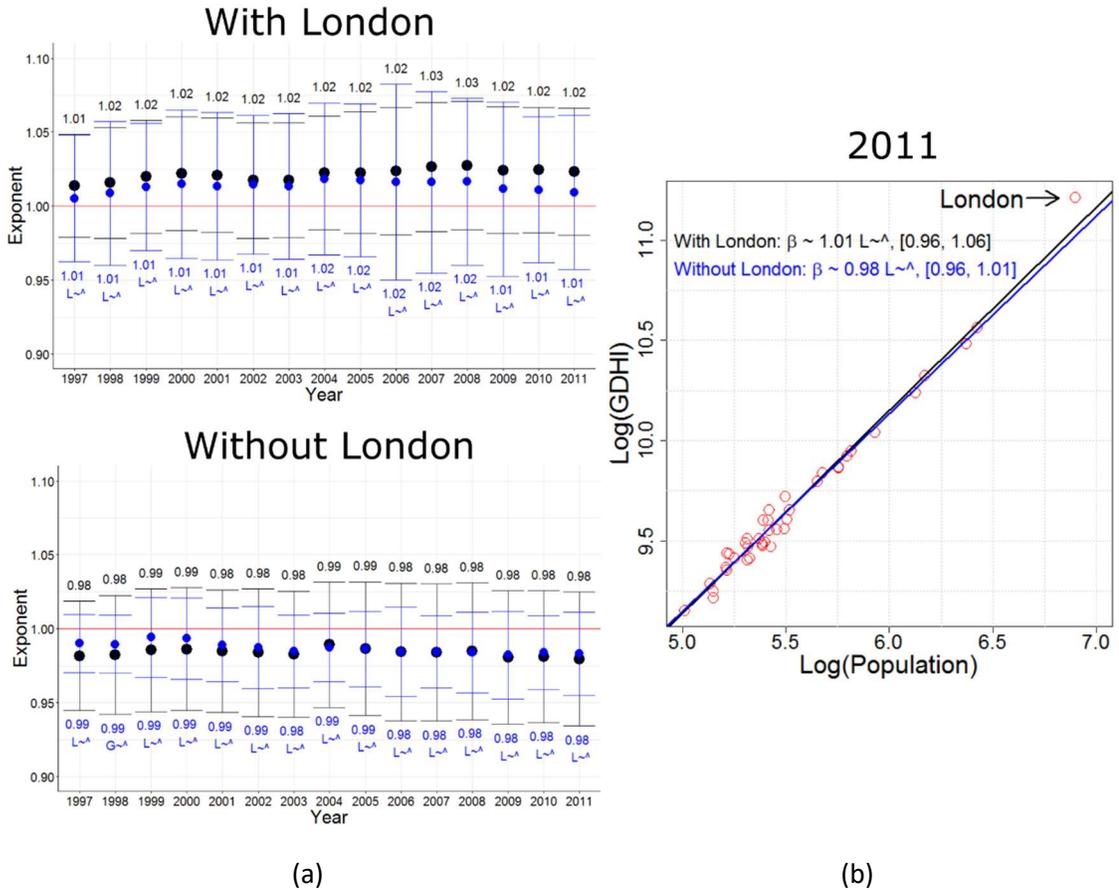

(a)  (b)

Figure S2: (a) The scaling exponents and 95% confidence intervals for GDHI with respect to city size constructed from NUTS 3 units for multiple years. London is included in the top graph but excluded from the bottom one; (b) the corresponding scatter plot for 2011.

The scaling exponents of the derived urban system for the period of 1997 to 2011 are shown in Figure S2a (top). We find that the relation is linear for all the investigated years. We also perform the analysis with the exclusion of the extreme value of London (Figure S2a, bottom). A scatterplot with two regression lines for the year of 2011 is shown in Figure S2b. The statistical model is not rejected in all cases.

We conclude that the relation between income and city size is linear.

### 6.2 Income in E&W by MSOA units 2012

Income estimates for 2012 for E&W were published for the level of Middle Layer Super Output Area (MSOA). To estimate the scaling coefficient for 2012, we aggregated the MSOA into clusters using the same method as was used in Section 6.1. We find that the relation is linear (Figure S3).

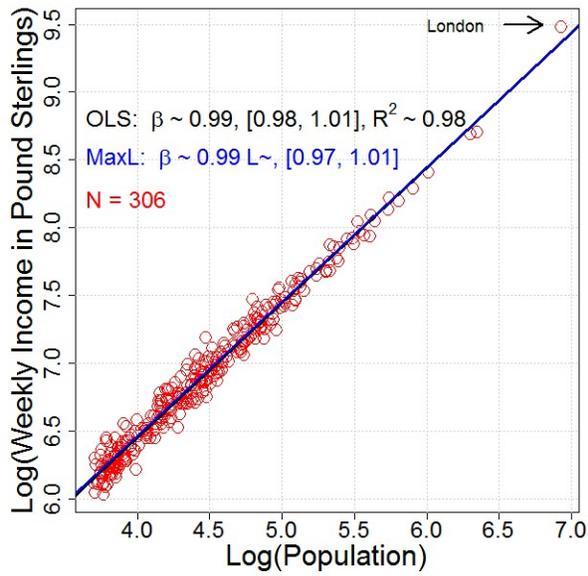

Figure S3: The scaling relation between total income and city size for cities of E&W in 2012.

### 6.3 The scaling exponents based on CBSA

Figures S4–S8 present all the paper's graphs for the US based on Core Based Statistical Areas (CBSAs). Census tract data were aggregated into 935 CBSAs.

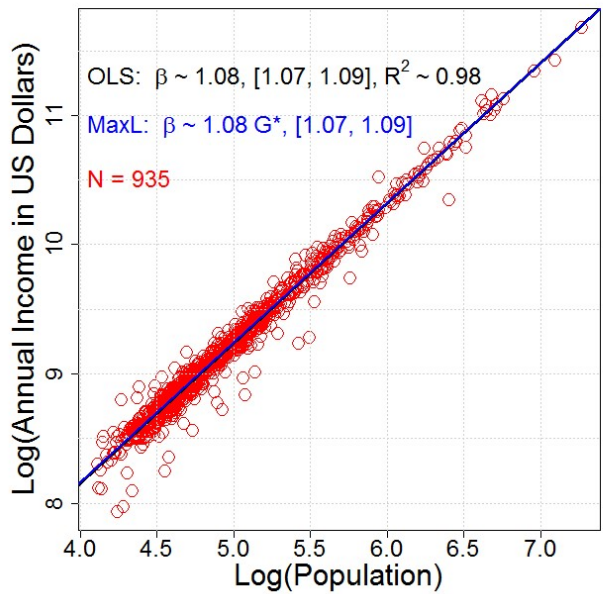

Figure S4: The scaling of total income by size for CBSAs (US)

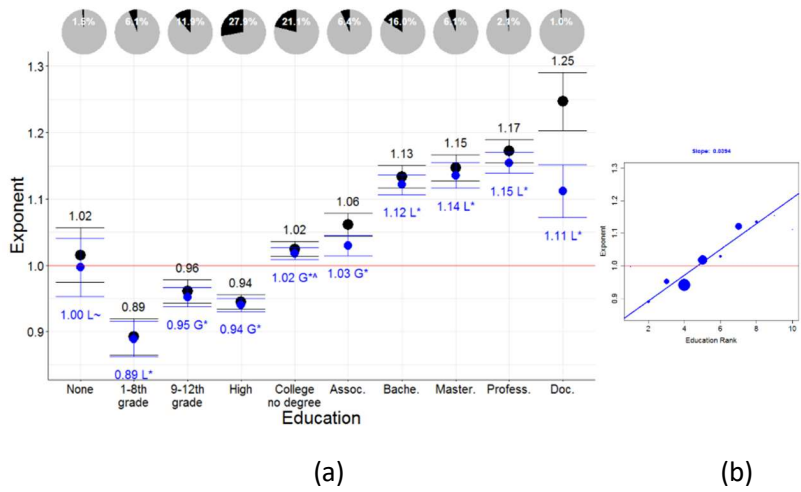

(a) (b)

Figure S5: (a) Scaling exponents for the highest level of education an individual completed in the US in 2000 based on CBSA. City size is defined as the number of adults (25 years and over). The categories are ordered from lower (left) to higher education levels (right). Pie charts depict the relative size of the educational cohort out of the total adult population. (b) The exponents versus the education rank. The size of the points is proportional to the size of the education cohort. The blue trend line is estimated using the weighted least squares method (the weight is the size of the cohort).

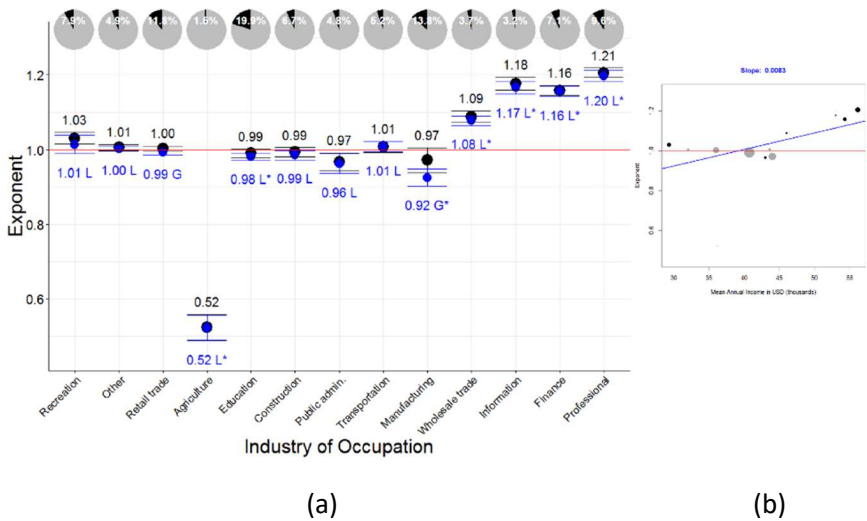

(a) (b)

Figure S6: (a) Scaling exponent for the number of adults (16 years and over) working in different industries in the US based on CBSA. City size is defined as the total number of employed adults. The industries are ordered according to national average pay from low (left) to high (right). Pie charts depict the fraction of individuals employed in each industry out of the total employed adult population. (b) The exponents versus the average pay. The size of the points is proportional to the size of the cohort. The blue trend line is estimated using the weighted least squares method (the weight is the size of the cohort).

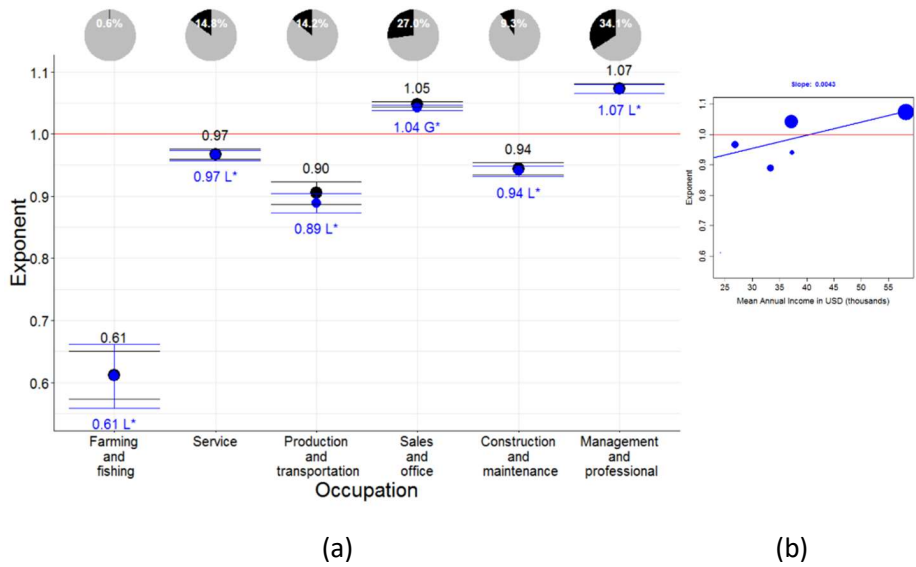

(a)    (b)

Figure S7: (a) Scaling exponent for the number of adults (16 years and over) working in different occupations in the US based on CBSA. City size is defined as the total number of employed adults. The employment categories are ordered according to national average pay from low (left) to high (right). Pie charts depict the fraction of individuals employed in a given occupation out of the total employed adult population. (b) The exponents versus the average pay. The size of the points is proportional to the size of the occupation cohort. The blue trend line is estimated using the weighted least squares method (the weight is the size of the cohort).

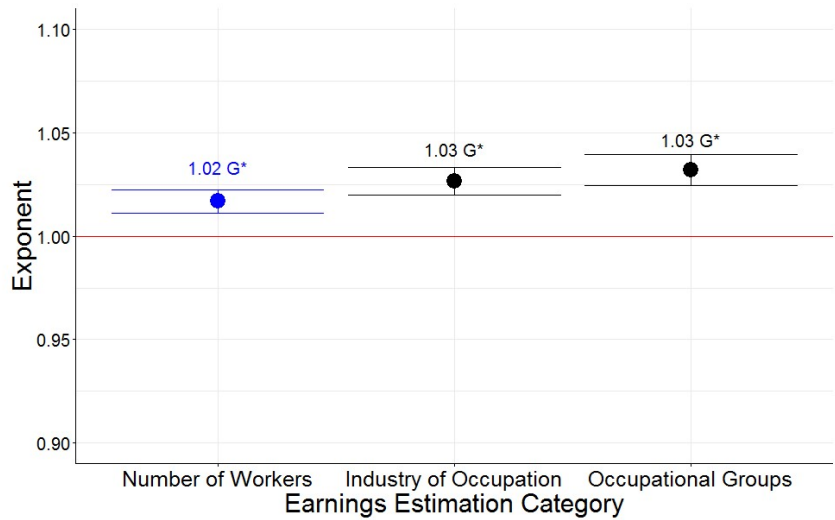

Figure S8: The scaling of estimated earnings in settlements in the US based on CBSA.

### 6.4. Data for England and Wales
#### 6.4.1. Ward units

The underlying spatial unit for all aggregations for E&W is the Census Area Statistics (CAS) ward definition produced by the UK Office for National Statistics. Ward boundaries reflect the political geography of the UK at a fine resolution and, due to the need to maintain equality of representation in political elections, have similar populations. CAS ward boundaries, in particular, have been the standard format for the release of ward-level census information since 2003. They reflect electoral ward boundaries promulgated as at 31 December 2002 and contain 8,850 separate wards for E&W. Census data used in the study were only available for England and Wales as the process used for collating data in Scotland and the definition of geographic boundaries meant that equivalent datasets could not be produced.

The data on population were taken from the UK census table "UV02".

### 6.4.2. Qualifications

Data of level of qualifications were taken from the Qualifications and Students (KS13) dataset. The highest level of qualification is derived from responses to both the qualifications question and the professional qualification question, and the levels attained relate to qualifications as follows:

1. No Qualifications: No academic, vocational, or professional qualifications.
2. Level 1: 1+ O level passes, 1+ CSE/GCSE any grades, NVQ level 1, Foundation GNVQ
3. Level 2: 5+ O level passes, 5+ CSEs (grade 1). 5+ GCSEs (grades A–C), School Certificate, 1+ A levels/AS levels, NVQ level 2, Intermediate GNVQ
4. Level 3: 2+ A levels, 4+ AS levels, Higher School certificate, NVQ level 3, Advanced GNVQ
5. Level 4/5: First degree, Higher degree, NVQ levels 4 and 5, HNC, HND, Qualified Teacher status, Qualified Medical Doctor, Qualified Dentist, Qualified Nurse, Midwife, Health Visitor

### 6.4.3. Industry of employment

Data on the industry of employment of employees were taken from the UK census "Industry of Employment (UV34)" table. The table shows the usual resident population aged 16 to 74 in employment by the industry in which individuals work. The industry in which a person works is determined by the response to the 2001 census question asking for a description of the business of the person's employer (or own business if self-employed). The responses were coded to a modified version of the UK Standard Industrial Classification of Economic Activities 1992 UK SIC (92). In the 2001 census, the industry of employment information was collected for usual residents. A usual resident was generally defined as someone who spent most of their time at a specific address. It included people who usually lived at that address but were temporarily away (on holiday, visiting friends or relatives, or temporarily in a hospital or similar establishment); people who worked away from home for part of the time; students, if it was their term-time address; a baby born before 30 April 2001 even if he or she was still in hospital; and people present on census day, even if temporarily, who had no other usual address. However, it did not include anyone present on census day who had another usual address or anyone who had been living or intended to live in a special establishment, such as a residential home, nursing home, or hospital, for six months or more.

Data on the national average gross weekly pay of each group were taken from Table 5 ("Region by Industry") of the 2001 Annual Survey of Hours and Earnings. The following table contains the industry of employment categories with their respective average earnings for E&W. In cases where the averages for E&W could not be calculated, the values for the entire UK were used instead:

| Industry of Employment | E&W: Average Weekly Earnings (GBP) | UK: Average Weekly Earnings (GBP) |
|---|---|---|
| Hotels and restaurants | 185.295 | 182.5 |
| Agriculture, hunting, and forestry | | 266.8 |
| Wholesale and retail trade, repairs | 302.537 | 297.6 |
| Health and social work | 309.601 | 309.8 |
| Education | 319.736 | 320.9 |
| Other community, social, and personal service activities | 338.968 | 333.0 |
| Public administration and defence, social security | 398.7094 | 395.6 |
| Transport, storage, and communications | 427.375 | 422.1 |
| Manufacturing | 427.123 | 422.4 |
| Construction | 432.817 | 427.7 |
| Real estate, renting, and business activities | 452.654 | 445.1 |
| Electricity, gas, and water supply | | 515.3 |
| Financial intermediation | 581.216 | 565.8 |
| Mining and quarrying | | 576.2 |
| Private households with employed persons | | 204.4 |
| Extra-territorial organizations and bodies | | 404.6 |

### 6.4.4. Occupational groups

Data on occupational groups were taken from the UK census table "KS12". The information on this table comes from responses to questions asking for the full title of the main job and a description of the job. The population includes any person aged 16 to 74 who carried out paid work in the week before the census, whether self-employed or an employee.

Data on the national average gross weekly pay of each group were taken from Table 15 ("Region by Occupation") of the 2001 Annual Survey of Hours and Earnings. The groups and averages are presented in the following table.

| Occupation | E&W: Average Weekly Earnings (GBD) | UK: Average Weekly Earnings (GBD) |
|---|---|---|
| Elementary occupations, (e.g., farm workers, laborers, kitchen assistants, and bar) | 193.1 | 191.3 |
| Sales and customer service | 205.7 | 203.1 |
| Personal service occupations (e.g., dental nurses, care assistants, nursery) | 213.6 | 215.9 |
| Administrative and secretarial occupations | 254 | 252.6 |
| Process, plant, and machine operatives | 326.7 | 325.3 |
| Skilled trades occupations (e.g., farmers, motor mechanics, electricians, plumbers) | 379.6 | 376.6 |
| Associate professional and technical occupations | 456.4 | 448.6 |
| Professional occupations (e.g., chemists, civil engineers, software professionals, dental practitioners, solicitors, and architects) | 552.3 | 549.8 |
| Managers and senior officials | 655.2 | 646.1 |

### 6.4.5. Income

The data on household income were taken from the UK census experimental statistics for 2001/02 and are provided at a fine geographic resolution for the whole of England and Wales. The original data and associated metadata for household income can be found under the "Topics" section of the UK Neighbourhood Statistics website (http://www.neighbourhood.statistics.gov.uk/). The income dataset corresponds to estimates that were produced using a model-based process which involves finding a relationship between survey data (data available on income) and other data drawn from administrative and census data sources. A model-fitting process was used to select covariates with a consistently strong relationship to the survey data. The strength of the relationship with these covariates was used to provide estimates on income for those wards where survey data on income were not available. Additional information on the provenance of the income data can be found on the Neighbourhood Statistics census website (www.neighbourhood.statistics.gov.uk).

In this study, we used the average weekly household total income (unequivalized) estimations in pounds sterling. We converted the average weekly income of each ward into a weekly sum by multiplying the average by the number of households. The number of households within each ward was taken from the Census "Household Composition" table ("UV65").

### 6.4.6. Commuting data

Data on commuting between ward of residence and ward of work were taken from the 2001 Census: Special Workplace Statistics (Levels 1, 2 & 3): Commuting Data at District, Ward and Output Area Scales dataset.

### 6.4.7. Land-cover maps

Land-cover data for E&W were taken from the 100 × 100 meter version of the Corine Land Cover 2000 raster data. The Corine Land Cover dataset is a pan-European land-use and land-cover mapping project that was initiated by the European Environment Agency. It supplies spatial data on the state of the European environmental landscape based on the interpretation of satellite imagery. It describes land cover (and partly land use) according to a nomenclature of 44 classes.

## 6.5. Data for the United States

The underlying spatial unit for all aggregations in the US is the census tract. These are small geographical units that partition the entire area of the US. They are produced by the US Census Bureau. They usually have a population size between 1,200 and 8,000 people, with an optimum size of 4,000 people (https://www.census.gov/geo/reference/webatlas/tracts.html). The Bureau creates census tracts to provide a stable set of boundaries for statistical comparison from census to census. Census tracts occasionally split due to population growth or merge when there is substantial population decline.

The GIS layers of the tracts were taken from the National Historical Geographic Information System (NHGIS) site (Minnesota Population Center, 2016).

### 6.5.1. Education

Educational attainment refers to the highest level of education that an individual has completed. This is distinct from the level of schooling that an individual has attended is attending. The data were taken from the "Sex by Educational Attainment for the Population 25 Years and Over (p037, sf3)" table of the 2000 census.

The data include the following levels of education: (1) no schooling completed; (2) nursery to 4th grade; (3) 5th and 6th grades; (4) 7th and 8th grades; (5) 9th grade; (6) 10th grade; (7) 11th grade; (8) 12th grade, no diploma; (9) high school graduate (includes equivalency); (10) some college, less than 1 year; (11) some college, 1 or more years, no degree; (12) associate degree; (13) bachelor's degree; (14) master's degree; (15) professional school degree; (16) doctoral degree.

### 6.5.2. Industry of occupation

The industry of occupation describes the kind of business conducted by a person's employing organization. These data are derived from a combination of write-in and check-box questions which are clerically coded by Census Bureau staff. The data were taken from the "Profile of Selected Economic Characteristics (DP-3, SF3)" table. The table contains the number of employed civilians aged 16 years and over working in each category.

Data on the average yearly earnings of each group were taken from Weinberg (2004). The following table contains the categories with their respective average earnings (from low to high).

| Industry of Occupation | Average Annual Earnings (USD) |
|---|---|
| Arts, entertainment, recreation, accommodation, and food services | 29,346.7 |
| Other services (except public administration) | 32,000.0 |
| Retail trade | 36,000.0 |

| Agriculture, forestry, fishing and hunting, and mining | 36,138.2 |
|---|---|
| Educational, health, and social services | 40,763.0 |
| Construction | 41,000.0 |
| Public administration | 43,000.0 |
| Transportation and warehousing, and utilities | 43,636.0 |
| Manufacturing | 44,000.0 |
| Wholesale trade | 46,000.0 |
| Information | 53,000.0 |
| Finance, insurance, real estate, and rental and leasing | 54,323.0 |
| Professional, scientific, management, administrative, and waste management services | 56,172.0 |

### 6.5.3. Occupational categories

Occupation describes the kind of work the person does on the job. Data on the number of individuals working in a given occupation were taken from the "Profile of Selected Economic Characteristics (DP-3, SF3)" table. These data are derived from responses to write-in questions that are clerically coded by Census Bureau staff. The dataset consists of the number of employed civilians aged 16 years and over who work in six occupational categories.

Data on average earnings was taken from Table 1 of the "Earnings Distribution of U.S. Year-Round Full-Time Workers by Occupation: 1999 (PHC-T-33)" dataset.

The occupation categories and average earnings are presented in the following table.

| Occupation | Average Annual Earnings (USD) |
|---|---|
| Farming, fishing, and forestry occupations | 24,191 |
| Service occupations | 26,885 |
| Production, transportation, and material moving occupations | 33,355 |
| Sales and office occupations | 37,121 |
| Construction, extraction, and maintenance occupations | 37,274 |
| Management, professional, and related occupations | 58,241 |

### 6.5.4. Income
Annual mean income of households and the number of households within each tract were taken from the "Income Distribution in 1999 of Households and Families: 2000 (QT-P32)" table. Total income was estimated by multiplying the two variables.

### 6.5.5. Commuting data
Commuting data were taken from the "Census 2000 Special Tabulation: Census Tract of Work by Census Tract of Residence (STP 64)" table. The table contains the number of workers in each tract by tract of residence commuter flow. Workers are defined as people aged 16 years and over who were employed and at work, full time or part time, during the census 2000 reference week. People who had a job but did not work during the reference week were excluded.

### 6.5.6. Land-cover maps

Land-cover data were taken from the National Land Cover Database of 2001 (NLCD2001) (Homer et al. 2007). The Land Cover Database is a classification scheme of 16 land-cover classes that has been applied consistently across all 50 United States and Puerto Rico at a spatial resolution of 30 meters. NLCD2001 is based primarily on the unsupervised classification of Landsat Enhanced Thematic Mapper+.

## 6.6. Clustering method

### 6.6.1. Density-based clustering

The density-based clustering method constructs clusters that represent cities and other settlements based on a density threshold. Each geographical unit (wards for E&W and census tracts for the US) is assigned a population density (persons per hectare). The procedure steps are outlined below along with a description of ArcMap implementation (Figure S9):

1. Selection of geographical units: Each unit that has a population density of $d$ persons per hectare or above is selected. This step is implemented using ArcMap's *Select (Analysis)* tool.
2. Creating geometrical features: adjacent units are merged into features. The geometries produced may not be polygons in the formal sense because they may contain holes. This step is done using ArcMaps's *Dissolve* tool.
3. Filling holes: In this step, we convert all the geometries into polygons to fill any holes. This is achieved using ArcMap's *FeatureToPolygon (Management)* tool.
4. Merging polygons by distance: Any pair of polygons that are no more than $m$ meters apart is merged into a single cluster. This step is calculated by creating a buffer (of distance $m/2$) around each polygon and merging polygons that intersect. This step is calculated using the *Buffer (Analysis)* and *Dissolve (Management)* tools.
5. Assigning a cluster membership to each geographical unit: We assign an identifier to each cluster. We then assign this identifier to each of the clusters constituting geographical units. This step is achieved by geographically joining the clusters and the geographical units using the *SpatialJoin (analysis)* tool.

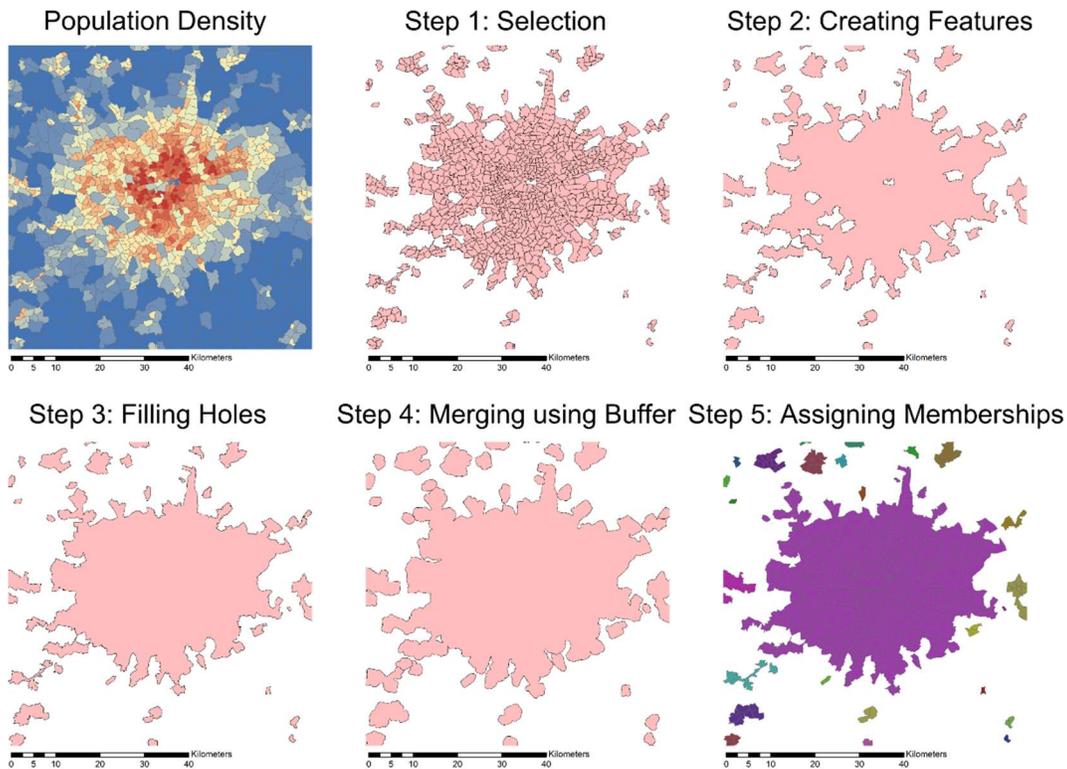

Figure S9: Illustration of the clustering method steps.

### 6.6.2. Incorporating commuting

We expand the density-based clusters by including areas from which workers commute:

1. For each geographical unit that does not belong to any cluster, we calculate the percentage of working residents who commute to each cluster.
2. We detect the cluster with the largest percentage (i.e., the cluster that receives that largest number of commuters from the given geographical unit). If two or more clusters share the same maximum percentage, we select one of these clusters with equal probability.
3. We assign the geographical unit to the selected cluster if the maximum percentage is larger than a *commuting threshold*.

For example, from a given geographical unit, 20% of the workers may commute to cluster 1 while 35% may commute to cluster 2. The rest may go to other clusters or to areas that are not part of any cluster. Let us assume that cluster 2 gets the largest percentage of commuters from the geographical units (35%) and the commuting threshold is 20%. In this case, we add the geographical units to cluster 2. However, if we set the commuting threshold to 40%, the geographical unit will not be added to any cluster. For low commuting thresholds, almost every geographical unit is assigned to a cluster, while for very high thresholds, almost all geographical units remain unassigned.

Figures 9 illustrate the urban systems constructed based on the density and commuting thresholds for the US and E&W.

### 6.6.3. Estimating the fit between clusters and land use

We estimate how well a set of clusters corresponds to the urban areas of a country by calculating the correlation between the clusters and a land-cover map (NLCD2001 for the US and the Corine dataset for E&W). We apply the following steps:

1. For each geographical unit (ward or census tract), we estimate the fraction of developed area out of the total area of the unit. We do so by calculating the fraction of raster cells with developed land cover out of the total number of cells that are within the geographical unit.
2. We assign a value of 1 to each geographic unit that is part of a cluster and 0 to a unit that is not part of any cluster.
3. Thus, each ward has a pair of values. We calculate the person product moment based on the pairs.

For the US, we consider the following land-cover categories of the NLCD2001 data as developed: developed low intensity (category 22), developed medium intensity (category 23), and developed high intensity (category 24).

For E&W, we considered the following land cover categories of the Corine data as developed: continuous urban fabric (category 1.1.1), discontinuous urban fabric (category 1.1.2), and green urban areas (category 1.4.1).

### 6.7. Moran's I estimation

The Moran's I index of autocorrelation between the average income (per capita) of clusters was calculated using the *SpatialAutocorrelation (Moran's I)* tool of ArcGIS 10.3. The spatial relationships between clusters were defined as the inverse square Euclidian distance. Only clusters that were within a given distance threshold (either 100, 200, or 300 kilometers) were considered.